\documentclass[12pt]{article}
\usepackage[authoryear,round]{natbib}
\usepackage{graphicx}
\usepackage{latexsym}
\usepackage[left=1in,top=1in,right=1in,bottom=1in]{geometry}\usepackage{amsmath}
\usepackage{booktabs}
\usepackage{color}
\usepackage{tikz}
\usetikzlibrary{shapes}
\usetikzlibrary{arrows}
\definecolor{darkblue}{rgb}{0,0.4,0.9}
\definecolor{gray10}{rgb}{0.1,0.1,0.1}
\definecolor{gray20}{rgb}{0.2,0.2,0.2}
\definecolor{gray30}{rgb}{0.3,0.3,0.3}
\definecolor{gray40}{rgb}{0.4,0.4,0.4}
\definecolor{gray60}{rgb}{0.6,0.6,0.6}
\definecolor{gray80}{rgb}{0.8,0.8,0.8}
\definecolor{gray90}{rgb}{0.9,0.9,.9}
\definecolor{gray95}{rgb}{0.95,0.95,.95}
\definecolor{gray96}{rgb}{0.96,0.96,.96}
\definecolor{lgreen} {RGB}{180,210,100}
\definecolor{dblue}  {RGB}{20,66,129}
\definecolor{ddblue} {RGB}{11,36,69}
\definecolor{lred}   {RGB}{220,0,0}
\definecolor{nred}   {RGB}{224,0,0}
\definecolor{norange}{RGB}{230,120,20}
\definecolor{nyellow}{RGB}{255,221,0}
\definecolor{ngreen} {RGB}{98,158,31}
\definecolor{dgreen} {RGB}{78,138,21}
\definecolor{nblue}  {RGB}{28,130,185}
\definecolor{jblue}  {RGB}{20,50,100}
\definecolor{nnyellow}{RGB}{235,200,0}
\definecolor{purple}{RGB}{150, 0, 120}
\definecolor{sgGreen} {RGB}{20, 180, 50}
\definecolor{revised}{rgb}{0,0,0.9}

\newtheorem{theorem}{Theorem}

\newcommand{\openr}{\hbox{${\rm I\kern-.2em R}$}}
\newcommand{\openn}{\hbox{${\rm I\kern-.2em N}$}}

\title{Technical Note: Targeted Maximum Likelihood Estimator for an ATE Standardized for New Target Population}\author{Mark J. van der Laan\textsuperscript{*}, Susan Gruber  \\  \small TL Revolution, LLC, Cambridge, MA \\ \\  \small \textsuperscript{*}Corresponding author: {\tt laan@TLrevolution.com}} \date{}

\begin{document}
\maketitle

\begin{abstract}
In this technical note we present a targeted maximum likelihood estimator (TMLE) for a previously studied target parameter that aims to transport an average treatment effect (ATE) on a clinical outcome in a source population to what the ATE would have been in another target population. It is assumed that one only observes baseline covariates in the target population, while we assume that one  can learn the conditional treatment effect on the outcome of interest in the source population. We also allow that one might observe only a subset of the covariates in the target population while all covariates are measured in the source population.  We consider the case that the outcome is a clinical outcome at some future time point that is subject to missingness, or that our outcome of interest is a time to event that is  subject to right-censoring.  We derive the canonical gradients and present the corresponding TMLEs for these two cases.

\end{abstract}

\section{Introduction}
When  accurate data on treatments and/or outcomes in a target population of interest are unavailable it is impossible to directly evaluate a causal effect of treatment.  However, if data on treatment, outcomes, and a set of baseline covariates that includes all confounders ($W$) are available in an external source population, when identifying assumptions are met the effect estimate can be transported to the target population of interest.  A causal transportability analysis is concerned with extending an estimated treatment effect in a source population to an external target population with possibly a different distribution of $W$.  

A recent review paper summarizes prior work in this area from the epidemiologic, economic, machine learning, and causal inference literature.$^1$ In this technical note we present a targeted maximum likelihood estimator (TMLE) for a previously studied target parameter that aims to transport an average treatment effect (ATE) on a clinical outcome in a source population to what the ATE would have been in another target population.$^{2,3}$ It is assumed that one only observes baseline covariates in the target population, while we assume that one  can learn the conditional treatment effect on the outcome of interest in the source population. We also allow that one might observe only a subset of the covariates in the target population while all covariates are measured in the source population.  We consider the case that the outcome is a clinical outcome at some future time point that is subject to missingness, or that our outcome of interest is a time to event that is  subject to right-censoring.  We derive the canonical gradients and present the corresponding TMLEs for these two cases.

\section{The statistical estimation problem}
We consider the following statistical estimation problem. 
Suppose we observe $n$ independent and identically distributed observations on a random variable $O=(S,S(W,A,\Delta,\Delta Y),(1-S)W)\sim P_0$, where $S$ is a binary indicator of belonging  to a source population on which we sample a complete data structure $(W,A,\Delta,\Delta Y)$ consisting of baseline covariates $W$, subsequent binary treatment $A$, a non-missing indicator $\Delta$ and outcome $Y$.
The sampled subjects with $S=0$ represents a sample from the target population of interest on which we only observe baseline covariates $W$.  
More generally, we consider the case that we only observe a partial set of covariates $V\subset W$ for the target population in which case our observed data structure is coded as $O=(S,S(W,A,\Delta, \Delta Y), (1-S)V)$. 
The likelihood of $O$ can be factorized as follows:
\begin{eqnarray*}
p(O)&=&p_S(S)p_V(V\mid S=0)^{1-S}p_{W}(W\mid S=1)^Sp_A(A\mid W,S=1)^S\\
&&p_{\Delta}(\Delta\mid W,A,S=1)^Sp_Y(Y\mid W,A,\Delta=1)^{\Delta S}.\end{eqnarray*}
We might assume models for the treatment mechanism $p_A$ and missingness mechanism $p_{\Delta}$ for the source population, but we make no further assumptions on the true density of the data $O$. Let ${\cal M}$ be the resulting nonparametric model for $P_0$. 
The statistical estimand is given by $\Psi:{\cal M}\rightarrow\openr$ defined by
\[
\Psi(P)=\Psi_1(P)-\Psi_0(P), \]
and
\[
\Psi_a(P)= \int_v dP(v\mid S=0) \int_w dP(w\mid V=v,S=1)E_P(Y\mid A=a,W=w,S=1).\]
Consider a structural causal model $S=f_S(U_S)$, $V=f_V(U_V)$ if $S=0$, $W=f_W(U_W)$ if $S=1$,
$A=f_A(W,U_A)$ if $S=1$, $\Delta=f_{\Delta}(W,A,U_{\Delta})$ if $S=1$, $Y=f_Y(W,A,U_Y)$ if $\Delta =1$ and $S=1$. This causal model allows us to define potential outcomes $Y_0$ and $Y_1$. Suppose that $E(Y_1-Y_0\mid W,S=1)=E(Y_1-Y_0\mid S=0)$;  $A,\Delta$ is independent of $(Y_0,Y_1)$, given $W$ and $S=1$; and 
 $P(W\mid V,S=0)=P(W\mid V,S=1)$. Then it follows that
 \begin{eqnarray*}
 E(Y_1-Y_0\mid S=0)&=&E\left( E\left(  \{E(Y\mid W,A=1,\Delta=1)\mid V,S=1\right)\mid S=0\right)\\
 &&-E\left( E\left(  \{E(Y\mid W,A=0,\Delta=1)\mid V,S=1\right)\mid S=0\right  )\\
 &=&\Psi_1(P)-\Psi_0(P),\end{eqnarray*}
 establishing identification of the ATE for the target population.  
 These are generally very strong assumptions. On the other hand, one might only assume that the $S=1$ study satisfies the randomization assumption on $(A,\Delta)$ above so that the conditional average treatment effect (CATE) $E(Y_1-Y_0\mid W,S=1)$ is identified by
 $E(Y\mid W,A=1,\Delta=1,S=1)-E(Y\mid W,A=0,\Delta=1,S=1)$. Then, we still have the following causal interpretation of $\Psi_1(P)-\Psi_0(P)$:
 \[
 \Psi_1(P)-\Psi_0(P)= E_{P^*} E_P(Y_1-Y_0\mid W,S=1),\]
 where $dP^*(w)=dP(w\mid S=1,V=v)dP(v\mid S=0)$ is a marginal distribution of $W$ implied by the conditional distribution of $W$, given $V$ for the $S=1$ population and the marginal distribution of $V$ for the $S=0$ population. 
 One can then still view the standardization as a way to shift the ATE from $S=1$ towards a population that resembles the target population. In particular, if $V=W$, then $dP^*=dP(W\mid S=0)$ is the actual covariate distribution for the target population. 
 
 In Section \ref{section2} we present the canonical gradient of $\Psi$ and a corresponding TMLE.
 In Section \ref{section3} we consider the observed data structure $(S,(1-S)V,S(W,A,\tilde{T}=\min(T,C),\Delta=I(\tilde{T}=T))$ and estimation of the causal effect of treatment on survival at a future time $t_0$ standardized as above with respect to the covariate distribution of the target population. In that section we will derive the canonical gradient and present  the resulting TMLE.

\section{TMLE of the transported causal effect of treatment on a clinical outcome subject to missingness}\label{section2}
\subsection{The efficient influence curve}
Recall the definition of $\Psi(P)=\Psi_1(P)-\Psi_0(P)=E_{V\mid S=0} E[E(Y\mid A=1,W,S=1)-E(Y\mid A=0,W,S=1)\mid V,S=1]$. Under randomization assumptions on $(A,\Delta)$, given $W,S=1$, and the appropriate positivity assumption, we have that $\Psi(P)= E_{V\mid S=0}E(Y_1-Y_0\mid V,S=1)$. As mentioned earlier, if one is also willing to assume that $E(Y_1-Y_0\mid V,S=1)=E(Y_1-Y_0\mid V,S=0)$ then $\Psi(P)=E(Y_1-Y_0\mid S=0)$ as desired. 

A plug-in estimator of $\psi_0=\Psi(P_0)$ involves obtaining an estimator $\bar{Q}_n(A,W,S=1)$ of $\bar{Q}_0(A,W,\Delta=1,S=1)=E_0(Y\mid A,W,\Delta=1,S=1)$ and  a subsequent regression of $\bar{Q}_n(1,W,S=1) -\bar{Q}_n(0,W,S=1)$ onto $V$ given $S=1$.
Let $\bar{Q}_0^r(a,V,S=1)=E_0( E_0(Y\mid A,W,\Delta=1,S=1)\mid V,S=1)$. Then the latter regression yields an estimator $\bar{Q}_n^r $ of $\bar{Q}_0^r$.  Finally, we take the empirical mean over $V$, given $S=0$.
So a plug-in estimator has the form
\[
\Psi({\bf P}_n)=E_{P_n}\left( \{\bar{Q}_n^r(1,V,S=1)-\bar{Q}_n^r(0,V,S=1)\}\mid S=0\right).\]

Let's determine the efficient influence curve $D_{\Psi_1(),P}$ of $\Psi_1$ at $P$.

\begin{description}
\item[$\mathbf{D_{\Psi_1(),V,P}:}$]The component of $P(V\mid S=0)$ of the EIC is given by:

\[
D_{\Psi_1(),V,P}= \frac{I(S=0)}{P(S=0)} (\bar{Q}(1,V,S=1)-\Psi_1(P)).\]

\item[$\mathbf{D_{\Psi_1(),Y,P}:}$]The component of $P(Y\mid A=1,W,\Delta=1,S=1)$ of the EIC is given by
\begin{align*}
D_{\Psi_1(),Y,P} &= C_{\psi_1,Y}\left(Y-\bar{Q}(1,W,\Delta=1,S=1)\right),
\end{align*}
where
\[
C_{\psi_1,Y}=\frac{I(A=1,\Delta=1,S=1)}{P(S=1)P(A=1\mid S=1,W)P(\Delta=1\mid S=1,W,A)} * \frac{P(V\mid S=0)}{P(V\mid S=1)} .
\]

Thus notice that we have $P(V\mid S=0)/P(V\mid S=1)$ instead of $P(W\mid S=0)/P(W\mid S=1)$ in the case that $V=W$. This factor can also be represented as:
\[
\frac{P(V\mid S=0)}{P(V\mid S=1)}=P(S=0\mid V)/P(S=1\mid V).\]

\item[$\mathbf{D_{\Psi_1(),W\mid V,P}:}$] Finally, the component of $P(W\mid S=1,V)$ of EIC is given by:
\begin{align*}
D_{\Psi_1(),W\mid V,P} &= \frac{P(V\mid S=0)}{P(V\mid S=1)} * \frac{I(S=1)}{P(S=1)} * \left(\bar{Q}(1,W,\Delta=1,S=1)-\bar{Q}^r(1,V,S=1)\right) \\
&= C_{\psi_1,W\mid V} * \left(\bar{Q}(1,W,\Delta=1,S=1)-\bar{Q}^r(1,V,S=1)\right)
\end{align*}
\end{description}
\begin{theorem}
Consider $\Psi_1:{\cal M}\rightarrow\openr$ defined above as \[
\Psi_1(P)=E_{V\mid S=0}E(E(Y\mid W,A=1,\Delta=1,S=1)\mid V,S=1).\]
Its canonical gradient at $P$ is given by:
\[
D^*_{\Psi_1(),P}=D_{\Psi_1(),V,P}+D_{\Psi_1(),Y,P}+D_{\Psi_1(),W\mid V,P},\]
where
\begin{eqnarray*}
D_{\Psi_1(),V,P}&=& \frac{I(S=0)}{P(S=0)} (\bar{Q}(1,V,S=1)-\Psi_1(P))\\
D_{\Psi_1(),Y,P} &=& C_{\psi_1,Y} \left(Y-\bar{Q}(1,W,\Delta=1,S=1)\right)\\
D_{\Psi_1(),W\mid V,P} &=& C_{\psi_1,W\mid V} * \left(\bar{Q}(1,W,\Delta=1,S=1)-\bar{Q}^r(1,V,S=1)\right).
\end{eqnarray*}
The definitions  of $C_{\psi_1,Y}$ and $C_{\psi_1,W\mid V}$ are given above.
\end{theorem}

When $V = W$ the components of $D^*$ are unchanged, but we see there is no contribution from the $D_{\Psi_1(),W\mid V,P}$ component  because 
$\bar{Q}(1,W,\Delta=1,S=1)-\bar{Q}^r(1,V,S=1) = 0$.

Analogue we obtain the expression for $D^*_{\Psi_0(),P}$ simply replacing $I(A=1)$ by $I(A=0)$; $\bar{Q}(1,W,\Delta=1,S=1)$ by $\bar{Q}(0,W,\Delta=1,S=1)$ and $\bar{Q}^r(1,V,S=1)$ by $\bar{Q}^r(0,V,S=1)$ in the above formulas. 

\subsection{TMLE}
\label{sec:TMLE}
The canonical gradient implies a corresponding TMLE of $\Psi_1(P_0)$ which we will present now.

{\bf Targeting outcome regression:}
First we run a regression $\bar{Q}_n(1,W,\Delta=1,S=1)$ of $Y$ on $W,A=1,S=1,\Delta=1$. We target it with clever covariate/weight $C_{\psi_1,Y},$ involving an estimate of $P(A\mid W,S=1)$, $P(\Delta=1\mid S=1,A=1,W)$, and $P(S=1\mid V)$. So externally we need an estimator of $P(A\mid W,S=1)$, $P(\Delta=1\mid S=1,A=1,W)$, and $P(S=1\mid V)$ giving the ratio $P(S=0\mid V)/P(S=1\mid V)$. Depending on whether $Y$ is binary or continuous we might use a logistic fluctuation with corresponding log-likelihood or linear fluctuation with squared error loss, analogue to the typical TMLE for the ATE based on $(W,A,\Delta, \Delta Y)$, This yields a targeted outcome regression  $\bar{Q}_n^*(1,W,\Delta=1,S=1).$

{\bf Targeting regression of outcome regression on $V,S=1$:}
Then, we run a regression of $\bar{Q}_n^*(1,W,\Delta=1,S=1)$ on $V$ given $S=1$. We then target that regression with a clever covariate (using least square loss and linear link) $C_{\psi_1,W\mid V},$ which involves our estimator of $P(S=1\mid V)$ again. That gives us a targeted $\bar{Q}_n^{r,*}(1,V,S=1)$ of $\bar{Q}_0^r(1,V,S=1)$.

{\bf Empirical mean covariate distribution of target population:}
Finally, we take the empirical mean over $V$, given $S=0$, giving us the final plug-in estimator $\psi_{1,n}^*=E_{P_n}\left( \bar{Q}_n^{r,*}(1,V,S=1)\mid S=0\right)$ of $\Psi_1(P_0)$.

Analogously we can compute the TMLE of $\Psi_0(P_0)$. The difference of the two TMLEs then yields a TMLE 
$\psi_{1,n}^*-\psi_{0,n}^*$ of $\Psi(P_0)$.

Alternatively, we can directly target $\Psi_1(P_0)-\Psi_0(P_0)$ directly  by using as clever covariate in the outcome regression  $C_{\psi_1-\psi_0,Y}=C_{\psi_1,Y}-C_{\psi_0,Y}$ to target $\bar{Q}_n(A,W,\Delta=1,S=1)$ resulting in the targeted $\bar{Q}_n^*(A,W,\Delta=1,S=1)$.  We then take the difference $\bar{Q}_n^*(1,W,\Delta=1,S=1)-\bar{Q}_n^*(0,W,\Delta=1,S=1)$ and as above run a regression on $V$ given $S=1$ with the same targeting step, and finally we take the empirical mean over $V$ given $S=0$. Both TMLEs are appropriate. 

\subsection{Double robustness of the TMLE and statistical inference}
The above TMLE is consistent if either $G_0=(p_{0,A},p_{0,\Delta},p_{0,S\mid V})$ is consistently estimated or if $Q_0=(\bar{Q}_0,\bar{Q}_0^r)$ is consistently estimated.  This follows from computing the exact remainder
$R_{\Psi()}(P,P_0)\equiv \Psi(P)-\Psi(P_0)+P_0 D^*_{\Psi(),P}$ demonstrating that $R_{\Psi()}(P,P_0)=0$ if either $Q=Q_0$ or $G=G_0$. The TMLE is asymptotically efficient if both are consistently estimated at the appropriate rate that would be achieved by a highly adaptive lasso (HAL) or super-learner using HAL in the library.$^{4,5}$ Statistical inference can be achieved with Wald type confidence intervals $\psi_n^*\pm 1.96 \sigma_n/n^{1/2}$ with $\sigma_n^2$ an estimator of the variance of $D^*_{\Psi(),P_0}$.

\section{TMLE of a transported ATE on a survival outcome subject to right censoring}\label{section3}
In this case our observed data structure is given by $O=(S,W,SA,S\tilde{T},S\Delta)\sim P_0$, where $\tilde{T}=\min(T,C)$ and $\Delta=I(T\leq C)$. For simplicity, we consider the case that one observes all the covariates $W$ for the target population, but the results below generalize the the more general case that one only observes a $V\subset W$ for the $S=0$ observations. Under a causal model we have $T=T_A$, $C=C_A$, and the underlying full data is $(S,W,S(T_0,T_1))$ for which we can define a causal estimand $\psi^F=P(T_1>t_0\mid S=0)-P(T_0>t_0\mid S=0)$ or a less ambitious causal estimand $\psi^{F,r}=E_{W\mid S=0}\{P(T_1>t_0\mid W,S=1)-P(T_0>t_0\mid W,S=1)\}$.
The latter is the average over the target population covariate distribution of the conditional treatment effect given $W$ for the source population. As stated in the previous section, the latter can be identified from the observed data when only assuming that the CATE is identified for the source population, while the earlier also relies on the CATE for the $S=1$ population to be equal to the CATE of the $S=0$ population. Regarding identification assumptions, we will at minimal assume that
$C$ is independent of $T$, given $S=1,W,A$, and $A$ is conditionally randomized, given $W,S=1$, so that 
the CATE $P(T_1>t_0\mid W,S=1)-P(T_0>t_0\mid W,S=1)$ is identified and thereby  $\psi^{F,r}=\Psi(P)$ below. 
Under more stringent assumptions one could then also obtain that $\psi^F=\Psi(P)$.  Either way, the statistical estimand is given by $\Psi_1(P)-\Psi_0(P)$ defined below so that these considerations are only relevant for obtaining a causal interpretation and for assessing a causal gap of this statistical estimand with respect to the causal estimand in a sensitivity analysis. 

As in the previous section, we consider a nonparametric statistical model  ${\cal M}$ with possible statistical model assumptions on the treatment mechanism $P(A=1\mid W,S=1)$; and the conditional hazard of censoring $\lambda_C(t\mid S=1,W,A)$.  
Let
\[
\Psi_a(P)=E_{W\mid S=0} S(t_0\mid W,A=a,S=1),\]
where $S(t_0\mid W,A=a,S=1)=P(T>t_0\mid W,A=a,S=1)=\prod_{s\leq t_0}(1-\lambda_T(s\mid W,A=a,S=1))$ and $\lambda_T(s\mid W,A=a,S=1)$ is the conditional failure time hazard.
Let $N(t)=I(\tilde{T}\leq t,\Delta=1,S=1)$, and $A_c(t)=I(\tilde{T}\leq t,\Delta=0,S=1)$. Let $\bar{H}(t)=(W,A,\bar{N}(t),\bar{A}_c(t))$ be the history for these two counting processes  with respect to which we define their intensities. 
These intensities of $N$ and $A_c$ are given by
\begin{eqnarray*}
\lambda(t\mid S=1,\bar{H}(t))&\equiv& E(dN(t)\mid S=1,\bar{H}(t))=I(S=1,\tilde{T}\geq t) \lambda_T(t\mid W,A,S=1)\\
\alpha(t\mid S=1,\bar{H}(t)))&\equiv& E(dA_c(t)\mid S=1,\bar{H}(t))=I(S=1,\tilde{T}\geq t)\lambda_C(t\mid W,A,S=1).
\end{eqnarray*}
Let $\tau$ be a final fixed endpoint representing maximal follow up time for all subjects in the $S=1$ sample.
The likelihood of $O=o$, equivalently $(S,W,A,bar{N},\bar{A}_c)=(s,w,a,\bar{n},\bar{a}_c)$, can be parametrized as follows:
\begin{eqnarray*}
p(o)&=&p_S(s)p_W(w\mid S=s)g_A(a\mid S=1,W)^s\prod_{t\leq \tau} \times \\
&&\lambda(t\mid S=1,\bar{H}(t))^{dN(t)s}(1-\lambda(t\mid S=1,\bar{H}(t))^{(1-dN(t))s}\times
\\
&&\prod_{t\leq \tau}\alpha(t\mid S=1,\bar{H}(t))^{dA_c(t)s}(1-\alpha(t\mid S=1,\bar{H}(t))^{(1-dA_c(t))s}.
\end{eqnarray*}
Let \[
g(o)=p_S(s)g_A(a\mid S=1,w)^s\prod_{t\leq \tau}\alpha^{s dA_c(t)}(1-\alpha)^{s (1-dA_c(t))}\]
represent the factors of the likelihood $p(o)$ corresponding with the treatment and censoring mechanism.
Let $q(o)$ be the remaining factors of the likelihood, so that $p(o)=q(o)g(o)$. 
So $q$ involves $p(W\mid S)$ and $\lambda(t\mid W,A,S=1)$.
 Note that $p$ is indexed by $(p_S,p_W,g_A,\lambda,\alpha)$. We could thus write our model as ${\cal M}=\{p=qg: g\in {\cal G}\}$
 for some model ${\cal G}$ on $g$ implied by models on $g_A$ and $\alpha$. 
We note that our target parameter depends on $P$ through $(p_{W\mid S=0},\lambda(\cdot\mid S=1,H(\cdot)))$ and thus only depends on $P$ through $q$.

\subsection{The canonical gradient of $\Psi_a()$.}
We consider the model ${\cal M}(g)$ in which $g$ is viewed as known. We will determine the canonical gradient  of $\Psi_a()$ at $p=q g$ in this model, which also equals the canonical gradient at $p=q g$ in our actual model ${\cal M}$. We find the canonical gradient  by starting with an initial gradient  $D_{\Psi_a(),P}$ of $\Psi_a()$ at $P$ restricted to model ${\cal M}(g)$. We then project this initial gradient on the tangent space of ${\cal M}(g)$ at $P$, giving the canonical gradient in model ${\cal M}(g)$ and thereby our desired canonical gradient for the model ${\cal M}$ as well.

{\bf Tangent space in model ${\cal M}(g)$:}
\begin{itemize}
\item $g$ does not generate a tangent space in this model.
\item The tangent space of $p_{W\mid S}$ is given by \[
T_{p_{W}}(P)=\{h(W,S)-E(h(W,S)\mid S): h\}.\]
The projection operator is given by:
\[
\Pi(D\mid T_{p_W}(P))=E (D(O)\mid W,S)-E (D(O)\mid S).\]
\item The tangent space of the conditional failure time hazard $\lambda$ at a given time point  $t=1,\ldots,\tau$, is given by:
\begin{eqnarray*}
T_{\lambda(t)}(P)&=&S\{  h(dN(t),H(t))-E(h(dN(t),H(t))\mid S=1,H(t)): h\}\\
&=&S\{h(H(t))(dN(t)-\lambda(t\mid S=1,H(t)): h\}.\end{eqnarray*}
The projection operator onto this sub-Hilbert space $T_{\lambda()}(P)$ of $L^2_0(P)$ is given by:
\[
\begin{array}{l}
\Pi(D\mid T_{\lambda(t)}(P))=\\
 S \left\{ I(\tilde{T}\geq t)E(D(O)\mid dN(t)=1,H(t),S=1)-E(D(O)\mid dN(t)=0,H(t),S=1)\right\}\\
 \hfill (dN(t)-\lambda(t\mid W,A,S=1)).\end{array}
 \]
\item The tangent space of $T_{\lambda}(P)=\sum_{t=1}^{\tau}T_{\lambda(t)}(P)$ is an orthogonal sum of the spaces $T_{\lambda(t)}(P)$ so that the projection operator onto $T_{\lambda}(P)$ is given by:
\[
\begin{array}{l}
\Pi(D\mid T_{\lambda}(P))=\\
S \sum_{t=1}^{\tau}I(\tilde{T}\geq t)\left\{ E(D(O)\mid dN(t)=1,H(t),S=1)-E(D(O)\mid dN(t)=0,H(t),S=1)\right\}\\
\hfill (dN(t)-\lambda(t\mid W,A,S=1)).\end{array}
\]
\item We conclude that $T_{{\cal M}(g)}(P)=T_{p_W}\oplus T_{\lambda}(P)$ and its projection operator is given by:
\[
\begin{array}{l}
\Pi(D\mid T_{{\cal M}(g)}(P))=
E(D(O)\mid W,S)-E(D(O)\mid S)+\\
S\sum_{t=1}^{\tau}I(\tilde{T}\geq t)\left\{  E(D(O)\mid dN(t)=1,H(t),S=1)-E(D(O)\mid dN(t)=0,H(t),S=1)\right\}\\
\hfill (dN(t)-\lambda(t\mid W,A,S=1) ).
\end{array}
\]
\end{itemize}

{\bf Initial gradient:}
We will now derive an initial gradient. We first propose an estimator whose influence curve is trivially determined. The latter influence curve is then our gradient. 
Consider the following inverse probability of censoring and treatment weighted (IPCTW) estimator:
\[
\psi_{n1}=\frac{1}{n}\sum_{i=1}^n \frac{p_n(W_i\mid S=0)}{p_n(W_i\mid S=1)}\frac{I(A_i=1,\tilde{T}_i>t_0,S_i=1,\Delta_i=1)}{p(S=1)g_A(1\mid W_i,S=1)
\prod_{s<\tilde{T}_i}(1-\alpha(s\mid S=1,W_i,A=1))},\]
where we act as if $W$ is discrete and $p_n(w\mid S=s)$ is the empirical proportion $P_n(W=w\mid S=s)$.
We note that $\psi_{n1}=\hat{\Psi}_w(P_n,p_{W,n})$ is linear in empirical mean $P_n$ and also depends on $p_{W,n}$ representing the empirical distributions $p_n(w\mid S=s)$ for $s=0$ and $s=1$.
The influence curve of $\hat{\Psi}_w(P_n,p_W)$ treating $p_W$ as known  is  trivially given by (analogue to influence curve of an empirical mean):
\[
D_{1,P}(O)=\frac{p(W\mid S=0)}{p(W\mid S=1)}\frac{I(A=1,\tilde{T}>t_0,S=1,\Delta=1)}{p(S=1)g_A(1\mid W,S=1)
\prod_{s<\tilde{T}}(1-\alpha(s\mid S=1,W,A=1))}-\Psi_1(P).\]
It remains to determine the influence curve $D_{W,P}$ of $\hat{\Psi}_w(P,p_{W,n})$. Even though the latter can also be straightforwardly derived,  we can avoid doing the latter because of  the following argument.

Consider  a plug-in NPMLE $\Psi_1(p_{W\mid S=0,n},\lambda_n)$ of $\Psi_1(P)=\Psi_1(p_{W\mid S=0},\lambda)$ acting as if $W$ is discrete so that an NPMLE exists. The influence curve of this NPMLE will be the efficient influence curve in our model ${\cal M}$ and will also imply the efficient influence curve for general distributions of $W$ (not  just discrete). This influence curve has two components $D_{1,W,P}^*$, the influence curve of 
$\Psi_1(p_{W\mid S=0,n},\lambda)-\Psi_1(p_{W\mid S=0},\lambda)$, and the influence curve  $D_{1,\lambda,P}^*$ of 
$\Psi_1(p_{W\mid S=0},\lambda_n)-\Psi_1(p_{W\mid S=0},\lambda)$. 
Let's consider the influence curve of the second $\Psi_1(p_{W\mid S=0},\lambda_n)$. This is the same as the efficient influence curve in the model ${\cal M}(g,p_{W\mid S})$ in which both $g$ and $p_{W\mid S}$ are known. In this latter model $D_{1,P}$ is actually a gradient of the IPTW estimator that uses the known $p(W\mid S=0)/p(W\mid S=1)$. Therefore, the canonical gradient in this model ${\cal M}(g,p_{W\mid S})$ is given by the projection of $D_{1,P}$ onto the tangent space of model ${\cal M}(g,P_{W\mid S})$ which is  $T_{\lambda}(P)$.
So we can conclude that  $D_{1,\lambda,P}^*=\Pi(D_{1,P}\mid T_{\lambda}(P))$. 
Therefore, if we separately determine $D_{1,W,P}^*$, then we can conclude that the desired canonical gradient is given by
\[
D^*_{1,P}=\Pi(D_{1,P}\mid T_{\lambda}(P))+ D_{1,W,P}^*.\]
The influence curve of $\Psi_1(p_{W\mid S=0,n},\lambda)$ corresponds with the influence curve of an empirical mean estimator of $E(h(W)\mid S=0)$ for a given $h$. The latter would be given by  $I(S=0)/P(S=0) (h(W)-E(h(W)\mid S=0))$.
Thus, we can conclude that
\[
D_{1,W,P}^*= \frac{I(S=0)}{P(S=0)}\left( S(t_0\mid A=1,W)-E(S(t_0\mid A=1,W)\mid S=0)\right).\]

So our remaining task is to determine the projection of the IPTCW $D_{1,P}$ onto $T_{\lambda}$.

{\bf $\Pi(D_{1,P}\mid T_{\lambda}(P))$:}
This projection involves for each $t$ taking a conditional expectation given $W,S=1,H(t)$. But given $W$, the factor $p(W\mid S=0)/p(W\mid S=1)$ is known and comes in front. So the projection on $T_{\lambda}(P)$ can be written as $S/P(S=1) R(W)$, where $R(W)\equiv P(W\mid S=0)/P(W\mid S=1)$, times the projection of \[
\tilde{D}_{1,P}=\frac{I(A=1,\tilde{T}>t_0,\Delta=1)}{g_A(1\mid W,S=1)
\prod_{s<\tilde{T}}(1-\alpha(s\mid S=1,W,A=1))}\] onto $T_{\lambda}(P)$. However, $\Pi(\tilde{D}_{1,P}\mid T_{\lambda}(P))$ is identical to the $T_{\lambda}$ component of the EIC for the target parameter $E_{W\mid S=1}S(t_0\mid A=1,W,S=1)$ which acts as if we only observe data from $P(O\mid S=1)$, which is already known and implemented in the survtmle R package.$^{6,7}$  So we obtain
\[
D_{1,\lambda,P}^*=\frac{S}{P(S=1)}R(W) \sum_{t=1}^{\tau} I(\tilde{T}\geq t)H_{1,g,\alpha,\lambda}(t,W,A)(dN(t)-\lambda(t\mid W,A,S=1)),\]
where 
\[
H_{1,g,\alpha,\lambda}(t,W,A)=-\frac{I(A=1)}{g_A(1\mid W,S=1)\bar{G}(t-\mid A=1,W)}\frac{S(t_0\mid A=1,W)}{S(t\mid A=1,W)}I(t\leq t_0).\]

\begin{theorem}
Let $R(W)=P(S=0\mid W)/P(S=1\mid W)$.
The efficient influence curve of $\Psi_1(P)$ on model ${\cal M}$ is given by:
\[
D^*_{1,P}(O)=D_{1,W,P}^*(W,S)+D_{1,\lambda,P}^*(O),\]
where
\begin{eqnarray*}
D_{1,W,P}^*&=& \frac{I(S=0)}{P(S=0)}\left( S(t_0\mid A=1,W,S=1)-E(S(t_0\mid A=1,W)\mid S=0)\right)\\
D^*_{1,\lambda,P}&=&\frac{S}{P(S=1)}R(W) \sum_{t=1}^{\tau} I(\tilde{T}\geq t)H_{1,g,\alpha,\lambda}(t,W,A)(dN(t)-\lambda(t\mid W,A,S=1))\\
H_{1,g,\alpha,\lambda}(t,W,A)&=&-\frac{I(A=1)}{g_A(1\mid W,S=1)\bar{G}(t-\mid A=1,W)}\frac{S(t_0\mid A=1,W)}{S(t\mid A=1,W)}I(t\leq t_0).\end{eqnarray*}
\end{theorem}

\subsection{TMLE}
Suppose we already have implemented a TMLE of $E_W S(t_0\mid A=1,W)$,  for example, as implemented in survtmle. Then we can use the same TMLE of the conditional failure time hazard $\lambda_0(t\mid A=1,W)$ but only use the $S=1$ data and multiply the clever covariate $H_{1,g_n,\alpha_n,\lambda_n}$ with $I(S=1)/P_n(S=1) R_n(W)$,   where $R_n$ is an estimator of $R_0$, and thus involves logistic regression of $S$ on $W$. One can also put this multiplication factor in the weight of the log likelihood loss. This yields a targeted conditional failure time hazard $\lambda_n^*(t\mid W,A,S=1)$ which maps into a corresponding conditional survival function $S_n^*(t_0\mid A=1,W,S=1)$. We then take the empirical mean of $S_n^*(t_0\mid A=1,W,S=1)$ given $W,S=0$. This represents the plug-in TMLE $\Psi_1(P_n^*)$ of $\Psi_1(P_0)$. Similarly, we can compute the TMLE of $\Psi_0(P_0)$. We can also simultaneously target $\Psi_0(P_0)$ and $\Psi_1(P_0)$ by targeting $\lambda_n$ with two of these clever covariates and again using the weight $R_n(W)$ in the loss. We could also directly target $\Psi_1(P_0)-\Psi_0(P_0)$ by using a single clever covariate $H_{1,g_n,\alpha_n,\lambda_n}-H_{0,g_n,\alpha_n,\lambda_n}$ again using weight $R_n(W)$ in the loss.

\subsection{Double robustness of the TMLE}
The above TMLE is consistent if either $G_0=(p_{0,A},\alpha_0,p_{S\mid W})$ is consistently estimated or if $\lambda_0$ is consistently estimated (where  the empirical distribution of $W$, given $S=0$, is always consistent).  This follows from computing the exact remainder
$R_{\Psi()}(P,P_0)\equiv \Psi(P)-\Psi(P_0)+P_0 D^*_{\Psi(),P}$ demonstrating that $R_{\Psi()}(P,P_0)=0$ if either $\lambda=\lambda_0$ or $G=G_0$. The TMLE is asymptotically efficient if both are consistently estimated at the appropriate rate that would be achieved by HAL or super-learner using HAL for  the conditional failure time hazard and conditional censoring hazard. Statistical inference can be achieved with Wald type confidence intervals $\psi_n^*\pm 1.96 \sigma_n/n^{1/2}$ with $\sigma_n^2$ an estimator of the variance of $D^*_{\Psi(),P_0}$.


\section{A less aggressive TMLE that targets the conditional treatment effect the same as the TMLE of the ATE for the source population}

We note that the above described TMLEs for our target estimand, the average of the conditional treatment effect over the target population covariate distribution, 
differs from the TMLE of the regular ATE of the source population (non-transported) by the way it targets  the conditional mean outcome or conditional failure time hazard. The difference is solely due to the factor $R(W)=P(S=0|W)/P(S=1|W)$ (or $R(V)=P(S=0|V)/P(S=1|V)$)  that enters in the clever covariate in the TMLE update step (and optionally may have been used to weight the initial outcome regression). That is to say, if we set $R(W)=1$, then both TMLEs target the outcome regression and conditional failure time hazard the same way.
This extra factor $R(W)$ targets the outcome regression/failure time hazard so that its fit at covariate values that are more prevalent in the target population than the source population are subjected to an extra bias reduction. This extra bias reductions comes at an increase in variance if $R(W)$ is not close to 1.  
For larger sample size, the actual TMLE is the preferred method but in finite samples both are competitive estimators with a slightly different bias variance trade-off.
Therefore, one might also decide to use the  regular TMLE that sets $R(W)=1$. This still yields a valid estimator and under appropriate conditions, such as when using the Highly Adaptive Lasso for estimation of the outcome regression or conditional failure time hazard, this TMLE will still be asymptotically as valid as the actual TMLE reported in this article. 
We recommend to report both types of TMLE in a data analysis, where one might see a wider confidence interval and a shift in the estimate for the real TMLE relative to the less aggressive TMLE.   
Although in this less aggressive TMLE the loss  for the initial outcome regression and  targeting step is optimized for the source population rather than weighted toward the target population, marginalization is still with respect to the covariate distribution in the target population.  

This less aggressive TMLE also has a key advantage  in that it can be used when observation-level source population data are not available to analysts with access to target population data.  This situation will arise when source population data cannot be shared due to privacy concerns, HIPAA restrictions, data licensing restrictions,  etc.  Evaluating the causal effect in the target population using the less aggressive TMLE requires sharing only the targeted model for the outcome (or the hazard of failure in a time-to-event analysis) trained on the source population data, which is then applied to data on the target population.  Valid inference (standard error estimates, p-values, confidence intervals) requires also sharing the contribution to the EIC  from each of the $n$ independent units of observation in the source population. This vector of continuous values contains no identifiable information, posing no risk from a privacy-preserving perspective.

\section*{References}
\begin{enumerate}
\item Degtiar I, Rose S. A review of generalizability and transportability. \emph{Annu. Rev. Stat. Appl.}. 2023 Mar 9;10(1):501-24.
\item van Der Laan MJ, Rubin D. Targeted maximum likelihood learning. \emph{Int. J. Biostat}. 2006 Dec 28;2(1).
\item Rudolph KE, van der Laan MJ. Robust estimation of encouragement-design intervention effects transported across sites. \emph{J R Stat Soc Series B Stat Methodol}. 2017 Nov;79(5):1509-1525. doi: 10.1111/rssb.12213. Epub 2016 Oct 31. PMID: 29375249; PMCID: PMC5784860.
\item Benkeser D, van der Laan M. The Highly Adaptive Lasso Estimator. \emph{Proc Int Conf Data Sci Adv Anal.} 2016;2016:689-696. doi: 10.1109/DSAA.2016.93. Epub 2016 Dec 26. PMID: 29094111; PMCID: PMC5662030.
\item van der Laan M. A generally efficient targeted minimum loss based estimator based on the highly adaptive lasso.  \emph{Int. J. Biostat}. 2017 Nov 1;13(2).
\item van der Laan MJ, Rose S. \emph{Targeted learning: causal inference for observational and experimental data}. New York: Springer; 2011.
\item Benkeser D, Hejazi N. survtmle: Targeted minimum loss-based
  estimation for survival analysis. doi:10.5281/zenodo.835868
  {https://doi.org/10.5281/zenodo.835868}, R package version
  1.1.3.9000.2, {https://github.com/benkeser/survtmle}.
\end{enumerate}

\section*{}
\bf{Acknowledgement: This work was funded in part by Sanofi. }
\end{document}